\documentclass[aps,nofootinbib,showpacs,preprintnumbers,amsmath,amssymb]{revtex4}

\def\be{\begin{equation}}
\def\ee{\end{equation}}
\def\ba{\begin{eqnarray}}
\def\ea{\end{eqnarray}}
\def \bea{\begin{eqnarray}}
\def \eea{\end{eqnarray}}

\usepackage{graphics}
\usepackage{graphicx}
\usepackage{dcolumn}
\usepackage{bm}
\usepackage{epsfig}
\usepackage{graphicx}
\usepackage{multirow}
\usepackage{dcolumn}
\usepackage{graphicx,epsfig}%
\usepackage{subfig}
\def \ee{\end{equation}}
\def \be{\begin{equation}}
\def \bea{\begin{eqnarray}}
\def \eea{\end{eqnarray}}

\preprint{}
\begin{document}

\title{Scale Setting for Self-consistent Backgrounds}

\author{Benjamin Koch$^*$, Paola Rioseco$^*$, and Carlos Contreras$^\dagger$}
 \affiliation{
$^*$Instituto de F\'{i}sica, \\Pontificia Universidad Cat\'{o}lica de Chile, \\
Av. Vicu\~{n}a Mackenna 4860, Santiago, Chile \\
$^\dagger$ Departamento de F\'{i}sica, Universidad T\'{e}cnica Federico Santa Mar\'{i}a;\\
Casilla 110-V, Valpara\'{i}so, Chile;\\
}
\date{\today}

\begin{abstract}
The quest for finding self-consistent background solutions
in quantum field theory is closely related to the way one decides
to set the renormalization scale $k$.
This freedom in the choice of the scale setting can lead
to ambiguities and conceptual inconsistencies such as the
non-conservation of the stress-energy tensor.
In this paper a setting for the ``scale-field''  is proposed
at the level of effective action, which avoids such
inconsistencies by construction.
The mechanism and its potential is exemplified for scalar $\phi^4$ theory and for
Einstein-Hilbert-Maxwell theory.
\end{abstract}

\maketitle

%
\section{Introduction}
\label{sec:Intro}
\begin{figure}[t]
\centering
\includegraphics[width=10cm]{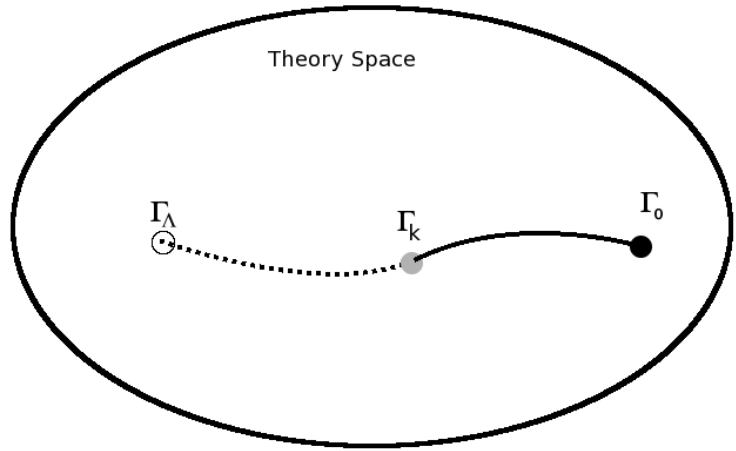}
\caption{\label{fig:Schema} Effective action $\Gamma_k$ in theory space where ideally $\Lambda \rightarrow \infty$
and the dotted line indicates an integration of $\phi_a$ over momentum degrees of freedom with $p\ge k$.}
\end{figure}

The effective action approach~\cite{Coleman:1973jx} can be seen as an elegant way of
defining a generating functional for
one-particle-irreducible Green's functions.
Following Wilson's idea~\cite{Wilson:1975}
one can study the effect of integrated quantum degrees
of freedom at different scales $k$. The scale dependent effective
action $\Gamma_k$ is to be understood as interpolation between the
ultraviolet (UV) bare action $\Gamma_\infty$ and the fully integrated action in the IR $\Gamma_0$
as it is sketched in figure~\ref{fig:Schema}.
$\Gamma_k$ contains scale dependent couplings $g^a_k$ which are obtained from
a suitable flow equation $k\partial_k \Gamma_k=\dots$~.
The space of solutions $g^a_k$ is called the ``coupling flow''~\cite{GellMann:1954fq}.
A specific trajectory is selected out of this flow by imposing conditions for the couplings at an initial scale $k_0$.
The evaluation of the effective action $\Gamma_k$ is typically hampered
by various technical difficulties such as singularities, anomalies, and non-localities.
However, in many cases those difficulties can be overcome by the ``regularization - renormalization'' technique,
where infinities are absorbed in the initial conditions at a scale $k_0$.
The technical details of this procedure will not be presented here, since they
are not relevant for the following discussion.
It will be assumed that the effective action $\Gamma_k$
has already been calculated.

Minimizing a given effective action with respect to variations of its (average) field content $\phi_a$
gives the equations of motion of
the effective action
\be\label{gap0}
\frac{\delta \Gamma_k}{\delta \phi_a}=0\quad.
\ee
Those equations are typically non-linear and sometimes non-local
differential equations and are frequently referred to as ``gap equations''~\cite{Avan:1983bv}.
Solutions of those equations have minimal energy (Gibbs free energy in
statistical mechanics). Therefore, finding solutions for the ``gap equations''
is highly relevant for defining a self-consistent background in quantum field theory.

However, even if it is technically possible to solve the ``gap equations'',
the physical interpretation of such a solution is still biased by the way the scale
$k$ is related to the quantities $x_i, \,Q_i, \dots$ (for example positions and charges)
that are used to describe the
physical system. Choosing a relation $k=k(x_i, \,Q_i, \dots)$ is called ``scale setting''.
The main focus of this article is on the role of scale setting in the quest of
finding self-consistent solutions of (\ref{gap0}).

The paper is organized as follows:
In section \ref{sec:Imp} the approach of improving classical solutions,
is studied and a criterion for scale setting is proposed.
Section \ref{sec:SSproposed} goes beyond improving classical solutions
by studying the ``gap equations''. A scale setting
is proposed in terms of an additional equation of motion.
The self-consistency and predictively of this approach is then studied
for specific examples
 such as scalar $\phi^4$ theory in section \ref{sec:phi4} and the
Einstein-Hilbert-Maxwell action in section \ref{sec:EHM}.
Conclusions can be found in section \ref{sec:sum}.

\section{A problem with scale setting in improving solutions schemes}
\label{sec:Imp}

The method of improving solutions has been successfully applied in many different contexts
\cite{Uehling:1935,Bonanno:1998ye,Bonanno:2000ep,Emoto:2005te,Bonanno:2006eu,Ward:2006vw,Falls:2010he,Basu:2010nf,Contreras:2013hua,Reuter:2006rg,Reuter:2010xb,Cai:2010zh,Frolov:2011rm,Falls:2012nd,Hewett:2007st,Litim:2007iu,Koch:2007yt,Burschil:2009va,Contillo:2011ag,Cai:2011kd, Hindmarsh:2011hx,Koch:2013owa,Litim:2013gga,Koch:2013rwa,Koch:2014cqa}.
This intuitive method is potentially useful when perturbation theory has limited reliability such
as for strong coupling or for non-renormalizable theories.

Instead of turning to one of the particular examples, this approach
will be discussed in quite generic fashion.
Lets assume a quite general $\phi^4$ quantum field theory with bare fields $\phi_i$ and generic couplings $\alpha, g$
with the equations of motion
\be\label{eomgen00}
 {\mathcal{D}} (\alpha \phi_a)+ \sum g^{abcd} (\phi_b\phi_c\phi_d)=0\quad,
\ee
where ${\mathcal{D}}$ is some differential operator (for example $\partial_\mu \partial^\mu$).
Lets further assume that the classical solution of (\ref{eomgen00}) is given by
\be
\phi_a=\phi^0_a(r,\alpha,g_n,A)\quad,
\ee
where $r$ are the coordinates and $A$ are integration constants.
Renormalization methods allow to calculate quantum corrections
to the couplings at arbitrary scale $k$, such that the couplings of the bare theory
become scale dependent quantities ($\alpha_k, g_k$).
How, do those supposedly small quantum corrections modify the form of the classical solution?
One approach to address this question is given the improving solutions procedure~\cite{Uehling:1935,Bonanno:1998ye},
where one assumes that the classical solution is actually the quantum solution at a certain scale say $k=k_0$.
For the case of simplicity lets choose $k_0=0$, 
just as it is done for the electromagnetic coupling in the standard model.
At this scale the classical equation of motion
\be\label{eomgen0}
 {\mathcal{D}} (\alpha_0 \phi^0_a)+ \sum g_{0}^{abcd} (\phi^0_b\phi^0_c\phi^0_d)=0\quad,
\ee
is solved by $\phi_a^0$
and the $k$ dependence is a small correction to this solution.
In this scheme one assumes that at first order, the functional form of the
solution stays unchanged, and only the couplings have to be replaced
by the scale dependent couplings in $\phi$
\be\label{improveAnsatz}
(\alpha_0,\, g_0)\rightarrow (\alpha_k,\, g_k)\quad.
\ee
Now one makes the ansatz that the classical solution obtains its first order
quantum correction only due to the scale dependence of the couplings
\bea\label{improved1}
\phi_a
&=& \phi_a^0(r,\alpha_0,g_0,A) + \frac{d}{dk}\phi_a(r,\alpha_k,g_k,A)|_{k=0} \cdot k +
{\mathcal{O}}(k^2)\quad,\\ \nonumber
\alpha_k&=&\alpha_0 +  \frac{d}{dk}\alpha_k |_{k=0} \cdot k +
{\mathcal{O}}(k^2)\quad,\\ \nonumber
g_k&=&g_0 +  \frac{d}{dk}g_k |_{k=0} \cdot k +
{\mathcal{O}}(k^2)\quad.
\eea
The second step is perform a scale setting which relates the
arbitrary scale $k$ to the physical coordinates
\be\label{scaleset0}
k \rightarrow k(r) \quad.
\ee
The explicit form for this scale setting is however a priory not uniquely determined
(in static spherically symmetric problems it has for example been proposed to use $k\sim 1/r$).
One first consistency check for this procedure would be to improve the equations of
motion (\ref{eomgen00}) in the same way and to check whether the improved
solution (\ref{improved1}) is actually a solution (up to order $k^1$) of those equations.
Inserting (\ref{improved1}) into (\ref{eomgen00}) one obtains
\bea\label{eomsecondline}
0&=&
{\mathcal{D}} (\alpha_0 \phi^0_a)+ \sum g_{0}^{abcd} (\phi^0_b\phi^0_c\phi^0_d)\\ \nonumber
&&+ {\mathcal{D}} \left[\left(k\frac{d}{dk}\right)(\alpha_k \phi_{a})|_{k=0}\right]+\left(k\frac{d}{dk}\right) \sum g_{k}^{abcd} (\phi^k_b\phi^k_c\phi^k_d)|_{k=0} + {\mathcal{O}}(k^2)\quad.
\eea
Using (\ref{eomgen0}) the first line is identically zero and one obtains that the second line has to be zero too,
if one wants to insist on the improved equations.
In most of the articles cited above it was not imposed that the improved equations of motion stay valid
and the scale setting was performed basically based on dimensional analysis.
At this point an important questions arises:
``Is the coupling $\alpha_k$ inside or outside the square brackets of the differential operator ${\mathcal{D}}$?''
This is a priory not clear since
starting from the equation of motion (\ref{eomgen0}) both alternatives would be equally possible.
However, this question can be ``answered'' (or say evaded) if one imposes a particular scale setting (\ref{scaleset0}) such that
the differential operator commutes with the scale function
\be\label{commuteSetting}
\left[{\mathcal{D}} , k(r)\right]=0\quad.
\ee
In this case the second line of (\ref{eomsecondline}) is equivalent to
\bea\label{eomsecondline2}
0&=&\left(k\frac{d}{dk}\right) \left.\left[{\mathcal{D}} \left[\phi^k_a)\right]
+ \sum \alpha_k^{-1} g_{k}^{abcd} (
\phi^k_b\phi^k_c\phi^k_d)\right]\right|_{k=0}
+ {\mathcal{O}}(k^2)\quad.
\eea
Thus, the remaining task would be first solving (\ref{commuteSetting}) for $k$
and then solving (\ref{eomsecondline2})
for $\phi_i^k$.

The fact that (\ref{commuteSetting}) can actually serve as a useful way of defining a scale setting
can be seen from a simple example:
For the three dimensional Laplace operator with spherical symmetry, the condition (\ref{commuteSetting}) reads
\be\label{laplaceSample}
\left[\frac{1}{r^2}\partial_r (r^2 \partial_r),\,\,k(r)\right]=0\quad.
\ee
It is solved for $r\neq 0$ by
\be
k(r)=\frac{\xi}{r}\quad,
\ee
which indeed agrees perfectly with the ad-hoc intuition coming from a dimensional analysis.

However, there is (at least) one other consistency condition one would like to impose,
the conservation of the stress-energy tensor, even at the quantum-improved level.
Let $T^0_{\mu \nu}$ be the classically conserved stress-energy tensor
\be\label{Tmunu0}
\nabla^\mu T^0_{\mu \nu}=0\quad,
\ee
then the straight forward improved stress-energy tensor would be taken to be
\be\label{Tmunuexp}
T_{\mu \nu}=T^0_{\mu \nu}+\left(\frac{d}{dk}T^k_{\mu \nu}|_{k=0}\right)\cdot k +{\mathcal{O}}(k^2)\quad.
\ee
Imposing conservation of (\ref{Tmunuexp}) and using (\ref{Tmunu0}) one finds
to leading order in $k$
\be\label{Tmunucond}
\nabla^\mu T_{\mu \nu}=
\nabla^\mu\left( \left(\frac{d}{dk}T^k_{\mu \nu}|_{k=0}\right)\cdot k\right)
\equiv 0\quad.
\ee
If one would try to solve this problem in the same spirit as (\ref{commuteSetting})
by imposing $[\nabla^\mu, k(r)]\equiv 0$ one easily finds that this is overly restrictive
allowing only for trivial solutions.
To circumvent this problem one can modify the definition  (\ref{improveAnsatz}) of the stress-energy
tensor (for example by using a different equation of state~\cite{Hindmarsh:2011hx})
but such an ad-hoc redefinition is not completely satisfactory.

To summarize, one sees that the method of improved solutions (\ref{eomsecondline})
can be made consistent, even at the level of improved equations of motions
if one imposes an adequate scale setting (\ref{commuteSetting})
(for example $k\sim 1/r$ for the spherical symmetric Laplacian).
However, it has limitations in the sense
that this procedure raises questions in the context of symmetries and conservation laws
and that it is restricted to first order corrections only.

This can be taken as motivation for seeking a more elegant way for obtaining a
description in the context of scale dependent couplings.

\section{The proposal:
Scale-field setting at the level of effective action}
\label{sec:SSproposed}

In the previous section it was shown how scale setting can be realized
in the improving solutions approach.
In this section a more general scale setting at the level of effective action will be proposed.

Lets assume that within the
quantum field theoretical model it was possible,
to evaluate the corresponding coupling flow and
to select a particular trajectory due to the choice of initial conditions ($g_i(k_0)=g_{i,0}$).
Thus, one can start with the
effective quantum action~\cite{Coleman:1973jx}
\be
\Gamma_k(\phi_a(x),\partial \phi_a(x),g^a_k)=\int d^4x \sqrt{-g} {\mathcal{L}}(\phi_a(x),\partial \phi_a(x),g^a_k)\quad,
\ee
where $\phi_a$ are actually the expectation values of the quantum fields
and $g^a_k$ are the scale dependent couplings,
including the coupling multiplying the kinetic term that is frequently
expressed in terms of field renormalization (see the first two subsections of \ref{sec:phi4}).

Note that doing this, one frequently has to truncate higher order-
or nonlocal couplings~\cite{Reuter:2002kd,Machado:2007ea,Codello:2010mj}
from the model, that might appear due to the quantum integration procedure.
In the following discussion it will however be assumed that all relevant couplings
are taken into account.
Now, one can derive  the equations of motion for the
average quantum fields $\phi_a$ from
\be
\frac{\delta \Gamma_k}{\delta \phi_a}=0\quad.
\ee
As mentioned in the introduction,
the solutions $\bar\phi_a(x,k)$ of those ``gap equations'' will also
be functions of the arbitrary scale $k$.
From a physical point of view this is however not yet satisfactory
since no possible observable can be a function of an a priory arbitrary scale.
In order to obtain a physical quantity one has to define some kind
of scale setting procedure, that establishes a relation between
the physical quantities (charges $Q_i$ and positions $x_j$) of a given problem and the
scale $k$. When doing this one can borrow an idea from
the calculation of observables $\langle T \phi(x_i)\dots \rangle_k$ in standard quantum field theory.
Also there, the observables turn out to be scale ``$k$'' dependent
quantities\footnote{It is, argued that this $k$ dependence is an artifact of the truncation in the loop expansion}.
Subsequently, the scale setting for those observables in terms of initial conditions and kinematical variables
$k=k(x_i;\;Q_i \dots)$ is chosen such that any $k$ dependence of the time ordered
correlation function (or some other observable) is minimized
\be\label{QFTcond0}
\left.\frac{d}{dk}\langle T \phi_1(x_1) \phi_2(x_2) \dots\rangle_k\right|_{k=k_{opt}}\equiv 0\quad.
\ee
This is the key philosophy that is used when deriving the 
``Callan-Symanzik'' equations~\cite{Callan:1970yg,Symanzik:1970rt},
the ``principal of minimal sensitivity~\cite{Stevenson:1981vj}, 
or the ``principle of maximal conformality''~\cite{Brodsky:1982gc,Brodsky:2011ig}.

It is proposed to implement an analogous philosophy at the level
of the effective action $\Gamma_k$.
This means that one should choose an optimal scale setting prescription for which
a variation of $k$ has a minimal impact on the self-consistent background $\bar \phi_i$.
This principle can be implemented by
promoting the a priory arbitrary scale to a physical scale-field in the effective quantum
action
\be\label{pres}
\Gamma_k(\phi_a(x),\partial \phi_a(x),g^a_k)\rightarrow \Gamma(\phi_a(x),\partial \phi_a(x),k(x),g^a_k)\quad.
\ee
This leads to the coupled equations of motion
\be\label{eomgen}
\frac{\delta \Gamma}{\delta \phi_a}=0\quad,\quad\left.
\frac{d}{dk}{\mathcal{L}}(\phi_a(x),\partial \phi_a(x),k(x),g^a_k)\right|_{k=k_{opt}}=0\quad.
\ee
Clearly it is not guaranteed that a solution for (\ref{eomgen}) can be found,
but such a prescription is not limited to be a variation of a classical solution
or to a saddle point approximation.
The procedure (\ref{eomgen}) has already been applied for some particular gravitational actions
\cite{Koch:2010nn,Domazet:2012tw,Davi:2013,Contreras:2013hua,Koch:2013rwa}
but in this work it is discussed in a broader context.
A nice feature of  such a procedure is that any solution of the equations (\ref{eomgen})
is automatically independent of $k$, which is actually the fundamental
precondition for a physical observable in the language of the renormalization group approach.

Promoting the scale $k$ to a scale-field $k(x)$ raises the question whether this new field 
only appears in the couplings $g_k^a$, or whether it has to
be equipped with other additional couplings, for instance a proper kinetic term.
A standard procedure when introducing new fields into a Lagrangian
is to incorporate actually all couplings that are in agreement with the symmetry
of the Lagrangian. This abundant freedom is then restricted by imposing some other additional conditions such as renormalizability,
simplicity, and/or agreement with experimental constraints.
However, in the presented approach the philosophy is different.
The scale-field $k$ is understood to have its origin in the process of renormalization and
throughout this process, no such extra couplings are taken into account.
In particular, the beta functions of the couplings $g_k^a$ are calculated without any additional couplings.
Therefore, the presented version of scale-field setting is chosen in a sense
``minimal'', since it contemplates the appearance of $k(x)$ only as dictated
by the running couplings $g_k^a$. 

Even though the prescription (\ref{pres}) looks quite convincing
from this perspective, it might be insufficient for example in the sense
that the space of solutions of (\ref{eomgen}) is actually empty,
apart from a trivial configuration or it is insufficient in the sense that the conservation of
the stress-energy tensor can not be guaranteed either.
Therefore, the idea will be studied for some examples,
where self-consistency of the approach
can be shown explicitly.

\section{Scale-field setting for scalar $\phi^4$ theory}
\label{sec:phi4}

As most simple example
without any further complications due to gauge symmetry lets
study the scale-field setting procedure for scalar $\phi^4$ theory.
There are various ways of writing the effective action
for $\phi^4$ theory. One of them is in terms of a scale dependent
wave function renormalization $Z_k$, running mass $m_k$, and running quartic coupling $g_k$.
The other way of writing this action is terms of separate couplings for every term appearing in the Lagrangian,
which are a coupling for the kinetic term $\alpha_k$, a coupling for the $\phi^2$ term $\tilde m^2_k$,
and a coupling for the quartic term $\tilde g_k$. As long as
the scale $k$ is assumed to be fixed, the formalism for both is
exactly equivalent. However, in the context of scale-field setting $k \rightarrow k(x)$,
derivatives do not necessarily commute with $k(x)$ 
 and both formulations could to be treated differently.
This subtlety will be exemplified in the following subsection, before
applying the scale setting to $\phi^4$ theory at the one loop level.

\subsection{Consistency in scalar $\phi^4$ theory}

If one works with separate couplings for every term appearing in the Lagrangian,
including the kinetic term,
 the effective action is
\be\label{scalarS}
\Gamma=\int d^4x \left( \frac{\alpha_k}{2}(\partial \phi)^2- \frac{\tilde m^2_k}{2}\phi^{2}-\frac{\tilde g_k}{4!}\phi^4\right)
\ee
with two fields $\phi$ and $k$.
The couplings $\alpha_k$, $\tilde m^2_k$, and $\tilde g_k$ are functions of the field $k$.
This implies an equation of motion for
 $\delta \phi$:
\be\label{eomphi}
\partial_\mu(\alpha_k \partial^\mu \phi )+ \tilde m^2_k \phi^{}+\frac{\tilde g_k}{6} \phi^3=0
\ee
and an other equation of motion for $k$:
\be\label{eomkphi}
\alpha_k' (\partial \phi)^2- (\tilde m^2_k)' \phi^{2}-\frac{1}{12} \tilde g_k' \phi^4=0\quad,
\ee
where $\alpha'=\partial_k \alpha= (\partial_x \alpha) dx/dk$.

The conserved energy momentum tensor is obtained as variation with respect to
the metric tensor
\be\label{tmunuphi}
T_{\mu \nu}=\alpha_k (\partial_\mu \phi)(\partial_\nu \phi)-
g_{\mu \nu}\left(
\frac{\alpha_k}{2}(\partial \phi)^2- \frac{\tilde m^2_k}{2}\phi^{2}-\frac{\tilde g_k}{4!}\phi^4
\right)\quad.
\ee
The corresponding conservation law reads
\bea \label{consT}
0&=&\partial^\mu T_{\mu \nu}\\ \nonumber
&=&
\partial^\mu (\alpha_k \partial_\mu \phi)\partial_\nu \phi + \alpha_k (\partial_\mu \phi) \partial^\mu \partial_\nu \phi
-\frac{1}{2}\partial_\nu
\left(\alpha_k(\partial \phi)^2- \tilde m^2_k \phi^2-\frac{\tilde g_k}{12}\phi^4
\right)\\ \nonumber
&=&-\frac{1}{2}\left(\alpha_k' (\partial \phi)^2- (\tilde m^2_k)' \phi^{2}-\frac{1}{12} \tilde g_k' \phi^4\right) \cdot \partial_\nu k\quad,
\eea
where in the second line the $\phi$ equation of motion (\ref{eomphi}) was used.
One easily observes that (\ref{consT}) is identical to the equation of motion for $k$
(\ref{eomkphi}).
This shows that the approach at the level of effective action (\ref{eomgen})
is on the one hand implementing the idea of
minimal scale dependence and on
the other hand maintaining the validity of improved equations of motion
and the fundamental conservation law (see section \ref{sec:Imp}).

Instead of writing a coupling for the kinetic term 
one frequently works with wave function renormalization 
where the bare field is $\phi_B=\sqrt{Z_k}\phi$.
In this case one could simply identify $\alpha_k=Z_k$, $\tilde m^2_k= Z_k m_k^2$, and $\tilde g_k=g_k Z_k^2$
and  observe that the corresponding effective action is completely equivalent to (\ref{scalarS}).
However, if one allows for field valued scales $k=k(x)$, this identification is not the only possibility,
since the derivatives of the kinetic term, acting on the scale field might contribute to the action.
Still, even in this case it can be shown that the scale setting procedure is 
consistent with the conservation law, just like in (\ref{consT}).
\subsection{Scale-field setting in the one loop expansion of $\phi^4$ theory}

The loop expansion of $\phi^4$ theory has been calculated up to
high order in perturbation theory~\cite{Kleinert:1991rg}.
For the following example will be restricted to the one-loop expansion of the beta functions (see~\cite{Peskin})
\bea
\gamma_{Z} & =&\frac{d \ln Z_k}{d \ln k^2}=0\quad,\\ \nonumber
\beta_g &=&\frac{d g_k}{d \ln k}=\frac{3}{16 \pi^2} g_k^2\quad, \\ \nonumber
\beta_{m^{2}}   &=&\frac{d  m_k^2}{d \ln k}= (- 2 +\frac{g_k}{16 \pi^2})m_k^2\quad.
\eea
Now one can integrate those flow equations
with initial conditions for $k=k_0$
\be\label{initialk0}
Z_{k_0}\equiv 1\,\quad{\mbox{and}} \quad g_{k_0}= g_0
\,\quad{\mbox{and}} \;\;m^2_{k_0}=m^2_0 \quad.
\ee
This determines the particular flow trajectory
\bea\label{scal1LoopCouplPesk}
Z_k&=&1\quad, \\ \nonumber
g_k&=& \frac{ g_0}{1-\frac{3}{16 \pi^2} g_0 \ln \left(k/k_0 \right)}\quad, \\ \nonumber
m^2_k&=&\frac{k_0^2}{ k^2} \frac{ m_0^2}{\left( 1 -\frac{3}{16\pi^2} g_0\log (k/k_0)\right)^{1/3}}=\frac{k_0^2 m_0^2}{ k^2} \left(\frac{ g_k}{g_0}\right)^{1/3}\quad.
\eea
In order to maintain legibility, the subscript of the scale dependent couplings
will be omitted ($m^2_k\rightarrow m^2$ and $g_k \rightarrow g$) in the following calculation.
Due to the constant wave function renormalization, the
equation of motion for $\phi$ simplifies to
\be\label{eomphisimpPesk}
\partial^2 \phi+m^2 \phi +\frac{g}{6} \phi^3=0
\ee
and the
 scale setting equation of motion (\ref{eomkphi}) simplifies to
\be\label{eomphikPesk}
\phi  ^3 g'
+12 \phi   (m^2)'
=0\quad.
\ee

\begin{figure}[t]
\centering
\includegraphics[width=10cm]{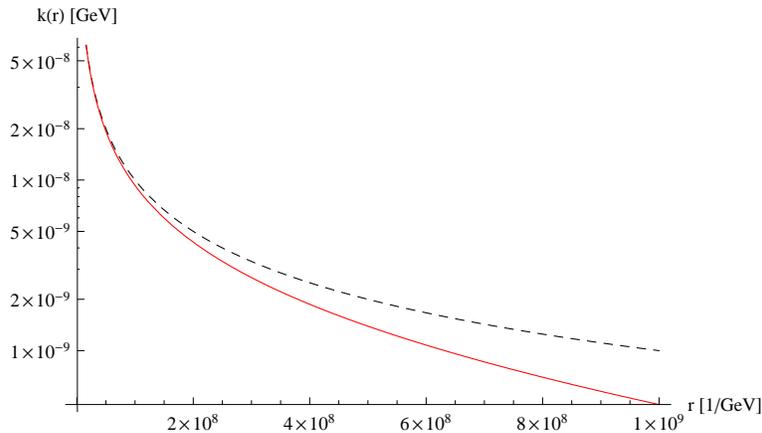}
\caption{\label{fig:kr}
Radial dependence of the scale setting (\ref{krsol}) in comparison to the
naive $1/r$ setting (dashed curve). The parameters used for the plot
are $m_0=1$~GeV, $g_0=0.5$, $c_1=1$, $c_2=0$, and $k_0=0$.}
\end{figure}

The two equations of motion (\ref{eomphisimpPesk}) and (\ref{eomphikPesk})
have to be solved for the two functions $k$ and $\phi$.
One way to approach this, is to first solve (\ref{eomphikPesk}) as non-differential
equation for $\phi^2$ giving
\be\label{phisol}
\phi^2=4 \frac{k_0^2 m_0^2}{k^2}\frac{  (32 \pi^2-g)}{  g^2}\left( \frac{g}{g_0}\right)^{(1/3)}\quad.
\ee
This can now be inserted into (\ref{eomphisimpPesk}) inducing a second order
differential equation for the scale setting, which after using the running couplings
(\ref{scal1LoopCouplPesk}) reads
\bea
A_k
\left[ \left(4(4 \pi)^8 + 11 (4 \pi)^6 g - \frac{(4 \pi)^4}{12} g^2+40 g^3-g^4\right)(\partial k )^2
- 8 \pi^2\left(\frac{(4 \pi)^6}{4}+ \frac{(4 \pi)^4}{6} g-96 \pi^2 g^2+g^3\right) k \cdot \partial^2 k
\right]&&\\ \nonumber
+\frac{2 k_0^3 m_0^2 (g+64 \pi^2)}{3 g^2 (g_0/g)^{1/3}k^3}
\sqrt{(m_0^2 (32 \pi^2-g))(g/g_0)^{1/3}}&=&0\quad,
\eea
where
\be
A_k=\frac{k_0}{64 \pi^2 g^{5/6}k^3 (-32 \pi^2 + g)^2}
\sqrt{\frac{m_0^2 (32 \pi^2-g)}{g_0^{1/3}}}\quad.
\ee
Within the validity of the beta functions (\ref{scal1LoopCouplPesk}),
which corresponds to first order in $g_0$, this equation simplifies
after the cancelation of a global factor to
\be\label{eomkdiffPesk}
k_0^2 m_0^2
\left(64 \pi^2 - 3 g_0 - 56 g_0 \ln (k/k_0) \right)
+ 6 g_0 (\partial k)^2 - 3 g_0 k \partial^2 k=0\quad .
\ee
The use of this relation, in terms of scale setting, will now be exemplified
for a specific system:

For static spherical symmetry the only allowed coordinate dependence
is with respect to the radial distance $k=k(r)$. In this case (\ref{eomkdiffPesk}) reads
\be\label{eomkdiffPeskRad}
k_0^2 m_0^2
\left(64 \pi^2 - 3 g_0 - 56 g_0 \ln (k/k_0) \right)
- 6 g_0 (\partial_r k)^2
+ \frac{3 g_0 k}{r^2}(2 r \partial_r k+ r^2 \partial_r^2 k)
=0\quad .
\ee
Due to the non-trivial structure of the differential operator
it is hard to find a general solution of (\ref{eomkdiffPeskRad}) but
one observes that, if $k_0 \neq 0$, this equation has actually a constant solution,
\be\label{kii}
k_i=k_0 \exp \left(-\frac{3}{56}+\frac{8 \pi^2}{7 g_0}\right)\quad.
\ee
From this expression one finds that demanding that $k_i=k_0 \neq 0$
would imply $g_0\approx 210$, which is clearly not in the validity range of the small $g_0$
approximation. Thus, it is safe to say that $k_i \neq k_0$.
One can now
 investigate the scale setting in the
vicinity of this constant scale $k_i$
\be\label{sepak}
k(r)=k_i +\delta k(r)+{\mathcal{O}}\left(\delta k^2 \right)\quad.
\ee
Inserting this in (\ref{eomkdiffPeskRad}) and expanding to first order in $\delta k$
one obtains a simpler differential equation for the radial dependence of the scale-field
\be
56 m_0^2 r \delta k-3 \frac{k_i^2}{k_0^2}\left( 2 \partial_r \delta k + r\partial_r^2 \delta k\right)=0\quad.
\ee
This differential equation can be directly solved by
\be
\delta k(r)= \exp \left(-2 \sqrt{\frac{14}{3}} \frac{k_0}{k_i} m_0 r \right)\frac{c_1}{r}
+\exp \left(+2 \sqrt{\frac{14}{3}} \frac{k_0}{k_i} m_0 r \right)\frac{c_2}{r}\quad,
\ee
where $c_1$ and $c_2$ are the constants of integration.
They have to be set by additional conditions, for example one might impose that
$\delta k$ does not diverge for large radii, which implies that $c_2 =0$.
The scale-field setting is then
\be\label{krsol}
k(r)=k_i + \exp \left(-2 \sqrt{\frac{14}{3}} \frac{k_0}{k_i} m_0 r \right)\cdot \frac{c_1}{r}\quad.
\ee
One notes that (\ref{krsol}) actually reproduces the standard $1/r$ behavior
for very small radii, but it has two additional features with respect to the naive guess.
The first difference consists in the constant factor, which might be suppressed for 
the case of a small scale $k_0$ in the initial conditions (\ref{initialk0}).
The second difference is an exponential
suppression factor which is controlled by the mass $m_0$ and by the value of $g_0$.
In figure \ref{fig:kr} this $r$ dependence of the scale setting is shown
in comparison to the usual setting $k\sim 1/r$.
The figure confirms the intuitive behavior for small $r$
and shows the exponential suppression for larger $r$.
Please note that the exercise was made in order to study 
the scale setting $k(r)$ and that in a more complete analysis, also
the stability and the finiteness of the corresponding
solution for $\phi(r)$ has to be studied.

\section{Scale-field setting for Einstein Hilbert Maxwell action}
\label{sec:EHM}
%

\subsection{Consistency in Einstein Hilbert Maxwell case}

In order to show the
consistency
of the proposed scale setting,
 with conservation laws in a
less-trivial example, one can study the approach for gravity coupled
to a $U(1)$ gauge field and to a cosmological constant.
Gravity is exemplary for a non-trivial field theory that is notoriously
perturbatively not renormalizable and the situation becomes even less favorable when it is coupled to matter.
Still, there exist non-perturbative methods that allow to
calculate effective actions and scale dependent couplings for this theory
\cite{Dou:1997fg,Souma:1999at,Reuter:2001ag,Lauscher:2001rz,Litim:2001up,Reuter:2003ca,Reuter:2004nx,Codello:2007bd,Machado:2007ea,Eichhorn:2009ah,Benedetti:2009rx,Niedermaier:2009zz,Harst:2011zx,Benedetti:2012dx,Dietz:2012ic,Nagy:2012rn,Christiansen:2012rx,Demmel:2012ub,Codello:2013wxa,Christiansen:2014raa}.

Therefore, it is reasonable to investigate the proposed scale setting procedure
in the context of a gravitational action coupled to matter.
As example for such a coupled gravitational system
the Einstein-Hilbert-Maxwell action will be discussed
\begin{eqnarray}
 \Gamma_k[g_{\mu \nu}, A_{\alpha}]&=&\int_{M}  d^4 x
 \sqrt{-g}\left(\frac{R-2\Lambda_{ k}}{16\pi G_{
k}}-\frac{1}{4 e_{k}^2}F_{\mu \nu}F^{\mu \nu}\right)
\quad, \label{EHaction}
\end{eqnarray}
where $R$ is the Ricci scalar and $F_{\mu \nu}=D_\mu A_\nu- D_\nu A_\mu$ is the
antisymmetric electromagnetic field strength tensor.
The scale dependent couplings are thus, Newtons coupling
$G_k$, the cosmological coupling $\Lambda_k$, and the electromagnetic
coupling $e_k$.
Please note that the flow of those couplings has been derived non-perturbatively in
\cite{Harst:2011zx}.
As in (\ref{pres}, \ref{eomgen})
the scale $k^2$ will be considered
as field without kinetic term.
The equations of motion for the metric field in (\ref{EHaction}) are
\begin{eqnarray}
 G_{\mu\nu}=-g_{\mu\nu}\Lambda_{k}-
\Delta t_{\mu \nu}+8 \frac{\pi G_k}{e_k^2} T_{\mu \nu}\quad,
\label{eomg}
\end{eqnarray}
where the possible coordinate dependence of $G_k$
induces an additional contribution to the stress-energy tensor
\cite{Reuter:2003ca}
\be
\Delta t_{\mu \nu}=G_{
k}\left(g_{\mu\nu}\Box-\nabla_\mu\nabla_\nu\right)\frac{1}{G_{k}}\quad.
\ee
Further, the stress-energy tensor for the electromagnetic part is given by
\be
T_{\mu \nu}=F_\nu^{\;\alpha} F_{\mu \alpha}- \frac{1}{4}g_{\mu \nu} F_{\mu \nu}F^{\mu \nu}\quad.
\ee
The equations of motion (Maxwell equations) for this $U(1)$ gauge field are
\be\label{eomA}
D_\mu\left( \frac{1}{e_k^2 }F^{\mu \nu}\right)=0 \quad,
\ee
and finally the equations of motion for the scale-field $k$ are
\be\label{eomk}
\left[R\nabla_\mu \left(\frac{1}{G_k}\right)-
2\nabla_\mu\left(\frac{\Lambda_k}{G_k}\right)
-\nabla_\mu \left( \frac{4 \pi}{ e_k^2}\right)F_{\alpha \beta}F^{\alpha \beta}\right]\cdot (\partial^\mu k)
=0 \quad.
\ee
The above equations of motion are complemented
by the relations corresponding to gauge invariance of the system.
For the case of diffeomorphism invariance one has
\be\label{diffeo}
\nabla^\mu G_{\mu \nu}=0
\ee
and for the internal $U(1)$ gauge symmetry
the corresponding equations are
\be\label{MaxwHom}
\nabla_{[ \mu} F_{\alpha \beta ]}=0\quad.
\ee
Please note that one has to work with (\ref{MaxwHom}) and not with
$\nabla_{[ \mu}e^{-1} F_{\alpha \beta ]}=0$ since
the factor $e^{-2}$ appears explicitly in the action (\ref{EHaction})
and not in the covariant derivatives.

The new ingredient due to the scale-field
is the equation (\ref{eomk}), therefore it is important
 to check whether this equation is actually non-trivial and consistent
with the equations (\ref{eomg} and \ref{eomA}).
The consistency can be shown by explicitly deriving (\ref{eomk}) from  (\ref{eomg} and \ref{eomA})
and further imposing that the gauge symmetries reflected by (\ref{diffeo} and \ref{MaxwHom}) are maintained.
Starting from (\ref{eomg}) one imposes (\ref{diffeo})
\begin{equation}
\nabla ^{\mu} G_{\mu\nu} =0= - g_{\mu\nu} \Lambda '_{k} \partial ^{\mu} K + \nabla ^{\mu} \Delta  t_{\mu\nu} +8\pi G '_{k} e_k^{-2} T_{\mu\nu} \partial ^{\mu} K  +8\pi G_{k} \nabla ^{\mu} \left(e_k^{-2} T_{\mu\nu} \right)
\end{equation}
using
\begin{equation}
\nabla ^{\mu} \left(\nabla _{\mu}\nabla _{\nu}-\square g_{\mu\nu} \right) \frac{1}{G_{k}}= R_{\mu\nu} \nabla ^{\mu} \frac{1}{G_{k}}
\end{equation}
one can factorize a $(\partial^\mu k)$ in the whole expression except of a single term
involving the electromagnetic stress-energy tensor
\be
0=\left[ - g_{\mu\nu} G_{k}\Lambda '_{k}  + G '_{k} \left(G_{\mu\nu} + \Lambda _{k} g_{\mu\nu} - 8\pi e_k^{-2}G_{k}T_{\mu\nu} \right)- R_{\mu\nu} G' _{k} +8\pi G '_{k} G_{k}  e_k^{-2} T_{\mu\nu} \right] (\partial ^{\mu} k)
 +8\pi {G_{k} }^{2}\nabla ^{\mu} \left( e_k^{-2} T_{\mu\nu} \right)\quad.
\ee
However, by using (\ref{MaxwHom}) and the antisymmetry of $F_{\mu \nu}$ one can show that $(\partial^\mu k)$ can also be factorized from this term
\be
\nabla ^{\mu} \left(e_k^{-2} T_{\mu\nu} \right) =\left[-\frac{g_{\mu\nu} }{4}  (e_k^{-2})' F_{\alpha \beta}  F^{\alpha \beta} \right]( \partial ^{\mu} k) \quad.
\ee
Thus, one has
\be
0=\left[ - g_{\mu\nu} G_{k}\Lambda '_{k}  + G '_{k} \left(G_{\mu\nu} + \Lambda _{k} g_{\mu\nu} - 8\pi e_k^{-2}G_{k}T_{\mu\nu} \right)- R_{\mu\nu} G' _{k} +8\pi G '_{k} G_{k}  e_k^{-2} T_{\mu\nu}
 -8\pi {G_{k} }^{2} \frac{g_{\mu\nu} }{4}   (e_k^{-2})' F_{\alpha \beta}  F^{\alpha \beta}\right] (\partial ^{\mu} k)\quad.
\ee
Since in this proof, there has no scale setting $k\rightarrow k(x)$ been applied yet,
one can choose any direction for the vector $(\partial ^{\mu} k)$ and still the above relation has to hold.
The only non-trivial way for this to happen is that the quantity in squared brackets has to vanish.
Tracing this quantity over its two indices one gets
\be
0= 2 \frac{ \Lambda '_{k}  }{G_{k}}+\frac{ G '_{k} \left(R- 2\Lambda _{k} \right)}{G_{k} ^{2}}+4\pi \left( (e_k^{-2})'   F^{2}   \right)\quad,
\ee
which is indeed identical to (\ref{eomk}).

Therefore, the equation of motion (\ref{eomk}) is indeed
consistent with the other equations of motion of the system  (\ref{eomg} and \ref{eomA}) in combination with the symmetry relations (\ref{diffeo} and \ref{MaxwHom}).
This consistency does not guarantee that the system has physically reasonable and non-trivial solutions.
But it confirms again that, even for gauge and gravitational systems,
the approach at the level of effective action (\ref{eomgen})
is on the one hand an elegant way of minimizing scale setting ambiguities
 and on the other hand maintaining the validity of improved equations of motion
and the fundamental conservation laws of the effective action (see section \ref{sec:Imp}).

Please note that
given the fact that the functional form of the scale dependent couplings $G_k$, $\Lambda_k$, and $e_k$
is in most known cases highly non-trivial~\cite{Harst:2011zx} one can hardly expect to obtain an analytical solution
of the equations (\ref{eomg} and \ref{eomA}).
However, in~\cite{Koch:tobe} it will be shown that with just one simplifying assumption
it is actually possible to find meaningful black hole solutions
for two different truncations of (\ref{EHaction}).

\subsection{Integrating out the scale-field:\\
UV scale-field setting for Einstein Hilbert Maxwell action in Asymptotic Safety}
\label{sec:BB}

Since in the proposed method, the scale $k$ is actually a scale field
and since fields can be integrated out of an effective action by solving their equations of motion
and inserting back into the action, $k(x)$ can be treated in the same way.
By this, the method (\ref{pres}) can actually be used in order
to construct a scale - independent effective action $\tilde \Gamma$, by integrating out $k(x)$ of the scale dependent action $\Gamma_k$.
In a general setting one proceeds by first solving the equation of motion for $k$ in momentum space
\be
\frac{\partial {\mathcal{L}}}{\partial k}=\partial_\mu \frac{\partial {\mathcal{L}}}{\partial k_{,\mu}}\quad,
\ee
giving
 $k=k(\phi)$.
This can be
inserted it back into the effective action which then gives
\be\label{intoutcond}
\Gamma(\phi, k)\rightarrow \Gamma(\phi,k(\phi))=\tilde \Gamma(\phi)\quad.
\ee
By this procedure one can ``integrate out'' the scale-field $k$ and one obtains
a new effective action $\tilde \Gamma(\phi)$ which is a function of $\phi$ only.
This new scale free effective action differs from the original effective action $\Gamma_k(\phi)$ in the sense
that it automatically contains a self-consistent scale setting.

\begin{figure}[t]
\centering
\includegraphics[width=10cm]{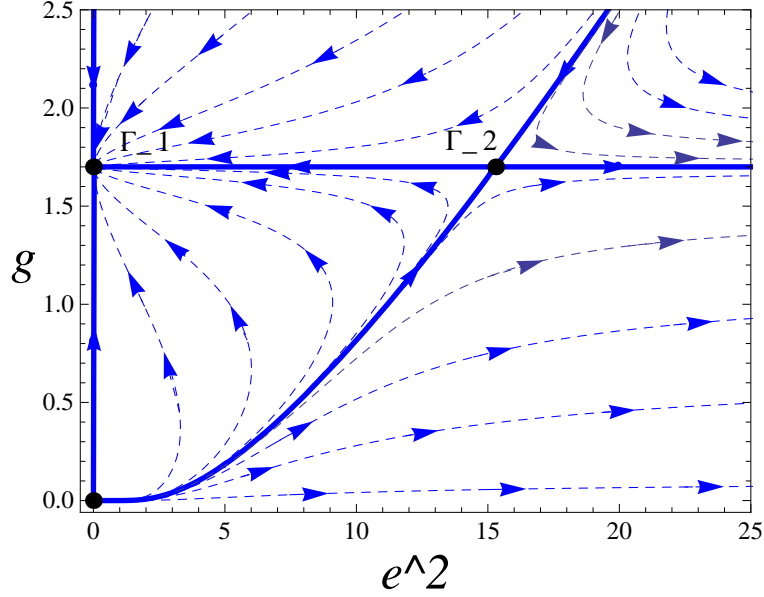}
\caption{\label{fig:eg}
Coupling flow projected to the dimensionless gravitational coupling $g$ and the electromagnetic coupling $e^2$.
The arrows point in direction of increasing $k$.}
\end{figure}

Lets exemplify this with the Einstein Hilbert Maxwell action (\ref{EHaction}).
In the deep UV limit ($k\rightarrow \infty$) there is strong evidence \cite{Souma:1999at,Reuter:2001ag,Litim:2001up}
for the existence of a non-Gaussian fixed point for the two gravitational couplings
\bea\label{UVGL}
G_k&\approx& \frac{g^*}{k^2}\quad,\\ \nonumber
\Lambda_k&\approx & \lambda^* k^2 \quad,
\eea
and there exists further evidence for (at least) one UV fixed point for the
electromagnetic couplings~\cite{Harst:2011zx}
\be\label{UVE}
\lim_{k\rightarrow \infty}\frac{1}{e_{k,2}^2}\approx \frac{1}{e^{*2}_2}\quad.
\ee
Since this fixed point in the electromagnetic coupling is not an attractor it is only approached
by particular trajectories in the corresponding flow.
Other trajectories either run into a Landau pole type of divergence at finite k,
or they run to vanishing values of $e_{k,1}$ at infinite $k$~\cite{Harst:2011zx}
\be \label{UVE1}
\lim_{k\rightarrow \infty}\frac{1}{e_{k,1}^2}\approx \frac{1}{e^{*2}_1}\cdot (k^2)^B\quad.
\ee
The value of the exponentiating factor $B$ for the second fixed point
depends on the method of calculation. Since the numerical value of $B$
ranges from $0.8$ to $1.6$~\cite{Harst:2011zx},
in this example the simplest possibility of $B\equiv1$ will be chosen.
The behavior of this flow projection using the functions of~\cite{Harst:2011zx}
is shown in figure \ref{fig:eg}.
One observes three fixed points in this flow:
Apart from the Gaussian fixed point at vanishing couplings
one sees, one fixed point with finite $g^*$ but vanishing $e^2$ (labeled $\Gamma_1$), which corresponds
to the limit (\ref{UVE1}), and an other fixed point with finite $g^*$ and $e^2$ (labeled $\Gamma_2$),
which corresponds to the limit (\ref{UVE}).
In order to integrate out the scale-field from the effective action (\ref{EHaction}) one has to solve the
corresponding equation of motion (\ref{eomk}) for $k^2$.
In the UV limit (\ref{UVGL}) one finds for the fixed point in (\ref{UVE})
\be\label{ss2}
k^2_2|_{UV}=\frac{R}{4 \lambda^*}\quad
\ee
and for the asymptotic behavior (\ref{UVE1}) one finds in the same limit
\be\label{ss1}
k^2_1|_{UV}=\frac{R-\frac{4 \pi g^*}{e^{*2}_1}F^2}{4 \lambda^*}\quad.
\ee
Those field-scale settings relate the UV scale $k^2$ proportional to the curvature scalar $R$,
in agreement to what is frequently intuited in the literature \cite{Frolov:2011ys,Bonanno:2012jy,Hindmarsh:2012rc,Copeland:2013vva,Koch:2014cqa}.
But the relations (\ref{ss2}, \ref{ss1}) go beyond this intuition since they also determine the constant proportionality factor 
and possible modifications due to the electromagnetic field strength.

Now follows the ``integrating out'' (\ref{intoutcond}), where the scale-field is eliminated from the original action.
The effective actions valid in the deep UV are obtained, from (\ref{EHaction}) after using again the approximation (\ref{UVGL}).
For the fixed point in (\ref{UVE}) and the corresponding scale setting (\ref{ss2}) this gives
\be
\tilde \Gamma_{UV,2}=\int d^4x \sqrt{-g} \left[\frac{R^2}{128\, \pi g^* \lambda^*}-\frac{F^2}{4 e^{*2}_2}\right]\quad.
\ee
For the asymptotic behavior (\ref{UVE1}) and the corresponding scale setting (\ref{ss1}) the UV effective action
results to be
\be\label{tildeGamma}
\tilde \Gamma_{UV,1}=\int d^4x \sqrt{-g} \left[\frac{\left(R-\frac{4 \pi g^*}{e^{*2}_1} F^2\right)^2}{128 \,\pi g^* \lambda^*}\right]\quad.
\ee
One observes that for vanishing $F^2=0$,
the $R^2$ dependence of the UV effective action
in Asymptotic Safe gravity, which is indeed renormalizable~\cite{Stelle:1976gc},
is recovered in agreement with other studies in the literature
\cite{Bonanno:2012jy,Hindmarsh:2012rc,Koch:2014cqa}.
However, in addition to this expectable result, 
the scale-field procedure in combination with integrating out the additional $k$-field dependence,
allowed to predict a generalization of those 
results to the coupling to a finite electromagnetic field strength $F^2\neq 0$.
Those UV results confirm that the fixed points (\ref{UVE}) and (\ref{UVE1})
correspond to different physical systems with different effective equations of motion for the background.
Therefore, the UV behavior of the self-consistent background, which is solution of one of those equations
of motion, can be largely different for different trajectories,
depending on the fixed point which is 
approached by a specific trajectory in the RG flow (see figure \ref{fig:eg}).

\section{Summary and conclusion}
\label{sec:sum}
In this paper  the problem of scale setting
in the context of finding self-consistent solutions of the effective action
- ``gap equations'' was discussed.
First, the procedure of improving solutions for
non-perturbative problems was reviewed by the use of a quite generic example.
It was shown that one might define a scale setting (\ref{commuteSetting})
that keeps consistency at the level of improved equations of motion,
but it was also shown that usually the conservation of the stress-energy
tensor can not be guaranteed throughout the improving solutions procedure.

Then, in section \ref{sec:SSproposed} a novel procedure for the scale setting
is proposed, by promoting the scale $k$ to a scale-field $k(x)$ at the level of
effective action $\Gamma$. This proposal is the essential idea of the presented work.

In order to demonstrate the functionality of the new procedure, the following sections
are devoted to discuss the approach for emblematic field theories.
As first example scalar $\phi^4$ theory is discussed and
the consistent conservation of the stress-energy tensor, even after the scale setting,
is shown explicitly. The self-consistency is shown for two common
forms of the $\phi^4$ effective action.
In order to complement this general result, by a more practical example,
an approximated self-consistent scale setting for spherically symmetric backgrounds
in $\phi^4$ theory is calculated at the one loop level.

In its generality the scale-field method is not limited to the simple scalar $\phi^4$ model.
Instead it is expected to work for a very broad class of theories.
For example, it also is meant to work in the context of much richer gauge theories.
This much broader applicability is exemplified by studying
the scale setting prescription in gravity coupled
to an electromagnetic stress-energy tensor, represented by Einstein-Hilbert-Maxwell theory.
It is explicitly shown that also in this example
the conservation of a generalized stress-energy tensor is guaranteed by the scale-field setting.
As application, the UV scale-field setting of Asymptotically Safe gravity coupled to an electromagnetic
field strength is calculated
and the scale independent effective action (valid in the UV) of this theory is derived by integrating
the scale-field out.
Finally, we would like to mention that the idea is not limited to background calculations,
but it should also be applicable to scale setting problems in scattering theory in the conventional Feynman sense.

\section*{Acknowledgements}
The work of B.K.\ was supported by proj.\ Fondecyt 1120360
and anillo Atlas Andino 10201.
C.C \ was supported by proj.\ Fondecyt 1120360 and DGIP-UTFSM grant 11.13.12.
The work of P.R. \ was supported by beca de doctorado Conicyt.
We acknowledge Frank Saueressig, Jorge Alfaro, Maximo Banados, Gorazd Cvetic, and Marco Aurelio Diaz for
valuable discussions. We further acknowledge the help of C. Gonzalez D\'iaz in producing 
figure \ref{fig:eg}.


\begin{thebibliography}{15}



\bibitem{Coleman:1973jx}
  S.~R.~Coleman and E.~J.~Weinberg,
  Phys.\ Rev.\ D {\bf 7}, 1888 (1973).

\bibitem{Wilson:1975}
 K.~G.~Wilson,
 Rev.\ Mod.\ Phys.\ {\bf 47}, 773 (1975).

\bibitem{GellMann:1954fq}
  M.~Gell-Mann and F.~E.~Low,
  Phys.\ Rev.\  {\bf 95}, 1300 (1954).


\bibitem{Avan:1983bv}
  J.~Avan and H.~J.~de Vega,
  Phys.\ Rev.\ D {\bf 29}, 2891 (1984).


%

\bibitem{Uehling:1935}
  E.A.~Uehling
  Phys.\ Rev. {\bf 48}, 55 (1935).


\bibitem{Bonanno:1998ye}
  A.~Bonanno and M.~Reuter,
  Phys.\ Rev.\  D {\bf 60}, 084011 (1999);
  gr-qc/9811026.

\bibitem{Bonanno:2000ep}
  A.~Bonanno and M.~Reuter,
  Phys.\ Rev.\  D {\bf 62}, 043008 (2000);
  hep-th/0002196.

\bibitem{Emoto:2005te}
  H.~Emoto,
  hep-th/0511075.

\bibitem{Bonanno:2006eu}
  A.~Bonanno and M.~Reuter,
  Phys.\ Rev.\ D {\bf 73}, 083005 (2006);
  hep-th/0602159.

\bibitem{Ward:2006vw}
  B.~F.~L.~Ward,
  Acta Phys.\ Polon.\ B {\bf 37}, 1967 (2006);
  hep-ph/0605054.

\bibitem{Falls:2010he}
  K.~Falls, D.~F.~Litim and A.~Raghuraman,
  Int.\ J.\ Mod.\ Phys.\ A {\bf 27}, 1250019 (2012);
  arXiv:1002.0260.


\bibitem{Basu:2010nf}
  S.~Basu and D.~Mattingly,
  Phys.\ Rev.\ D {\bf 82}, 124017 (2010);
  arXiv:1006.0718.

\bibitem{Contreras:2013hua}
  C.~Contreras, B.~Koch and P.~Rioseco,
  Class.\ Quant.\ Grav.\  {\bf 30}, 175009 (2013)
  [arXiv:1303.3892 [astro-ph.CO]].


\bibitem{Reuter:2006rg}
  M.~Reuter and E.~Tuiran,
  hep-th/0612037.

\bibitem{Reuter:2010xb}
  M.~Reuter and E.~Tuiran,
  Phys.\ Rev.\ D {\bf 83}, 044041 (2011);
  arXiv:1009.3528.


\bibitem{Cai:2010zh}
  Y.-F.~Cai and D.~A.~Easson,
  JCAP {\bf 1009}, 002 (2010);
  arXiv:1007.1317.





\bibitem{Frolov:2011rm}
  A.~M.~Frolov and D.~M.~Wardlaw,
  European Physical Journal B 85, 348 - 351 (2012)
  [arXiv:1110.3433 [nucl-th]];
  A.~M.~Frolov,
  arXiv:1111.2303 [math-ph].

\bibitem{Falls:2012nd}
  K.~Falls and D.~F.~Litim,
  Phys.\ Rev.\ D {\bf 89}, 084002 (2014)
  [arXiv:1212.1821 [gr-qc]].



\bibitem{Hewett:2007st}
  J.~Hewett and T.~Rizzo,
  JHEP {\bf 0712}, 009 (2007);
  arXiv:0707.3182.

\bibitem{Litim:2007iu}
  D.~F.~Litim and T.~Plehn,
  Phys.\ Rev.\ Lett.\  {\bf 100}, 131301 (2008);
  arXiv:0707.3983.

\bibitem{Koch:2007yt}
  B.~Koch,
  Phys.\ Lett.\  B {\bf 663}, 334 (2008);
  arXiv:0707.4644.


\bibitem{Burschil:2009va}
  T.~Burschil and B.~Koch,
  Zh.\ Eksp.\ Teor.\ Fiz.\  {\bf 92}, 219 (2010);
  arXiv:0912.4517.



\bibitem{Contillo:2011ag}
  A.~Contillo, M.~Hindmarsh and C.~Rahmede,
  Phys.\ Rev.\ D {\bf 85}, 043501 (2012)
  [arXiv:1108.0422 [gr-qc]].

\bibitem{Cai:2011kd}
  Y.~-F.~Cai and D.~A.~Easson,
  Phys.\ Rev.\ D {\bf 84}, 103502 (2011)
  [arXiv:1107.5815 [hep-th]].

\bibitem{Hindmarsh:2011hx}
  M.~Hindmarsh, D.~Litim and C.~Rahmede,
  JCAP {\bf 1107}, 019 (2011)
  [arXiv:1101.5401 [gr-qc]].

\bibitem{Koch:2013owa}
  B.~Koch and F.~Saueressig,
  Class.\ Quant.\ Grav.\  {\bf 31}, 015006 (2014)
  [arXiv:1306.1546 [hep-th]].


\bibitem{Litim:2013gga}
  D.~F.~Litim and K.~Nikolakopoulos,
  JHEP {\bf 1404}, 021 (2014)
  [arXiv:1308.5630 [hep-th]].

\bibitem{Koch:2013rwa}
  B.~Koch, C.~Contreras, P.~Rioseco and F.~Saueressig,
  arXiv:1311.1121 [hep-th].

\bibitem{Koch:2014cqa}
  B.~Koch and F.~Saueressig,
  Int.\ J.\ Mod.\ Phys.\ A {\bf 29}, no. 8, 1430011 (2014)
  [arXiv:1401.4452 [hep-th]].




\bibitem{Reuter:2002kd}
  M.~Reuter and F.~Saueressig,
  Phys.\ Rev.\ D {\bf 66}, 125001 (2002)
  [hep-th/0206145].


\bibitem{Machado:2007ea}
  P.~F.~Machado and F.~Saueressig,
    Phys.\ Rev.\ D {\bf 77}, 124045 (2008),
  arXiv:0712.0445.


\bibitem{Codello:2010mj}
  A.~Codello,
  Annals Phys.\  {\bf 325}, 1727 (2010)
  [arXiv:1004.2171 [hep-th]].


\bibitem{Callan:1970yg}
  C.~G.~Callan, Jr.,
  Phys.\ Rev.\ D {\bf 2}, 1541 (1970).

\bibitem{Symanzik:1970rt}
  K.~Symanzik,
  Commun.\ Math.\ Phys.\  {\bf 18}, 227 (1970).

\bibitem{Stevenson:1981vj} 
  P.~M.~Stevenson,
  Phys.\ Rev.\ D {\bf 23}, 2916 (1981).

\bibitem{Brodsky:1982gc}
  S.~J.~Brodsky, G.~P.~Lepage and P.~B.~Mackenzie,
  Phys.\ Rev.\ D {\bf 28}, 228 (1983).

\bibitem{Brodsky:2011ig}
  S.~J.~Brodsky and L.~Di Giustino,
  Phys.\ Rev.\ D {\bf 86}, 085026 (2012)
  [arXiv:1107.0338 [hep-ph]].




\bibitem{Koch:2010nn}
  B.~Koch and I.~Ramirez,
  Class.\ Quant.\ Grav.\  {\bf 28}, 055008 (2011)
  [arXiv:1010.2799 [gr-qc]].




\bibitem{Domazet:2012tw}
  S.~Domazet and H.~Stefancic,
  Class.\ Quant.\ Grav.\  {\bf 29}, 235005 (2012)
  [arXiv:1204.1483 [gr-qc]].


%
\bibitem{Davi:2013}
Davi C. Rodrigues, Benjamin Koch, Oliver F. Piattella, and Ilya L. Shapiro
Proceedings 5 Verao Quantico, 2014.

\bibitem{Kleinert:1991rg}
  H.~Kleinert, J.~Neu, V.~Schulte-Frohlinde, K.~G.~Chetyrkin and S.~A.~Larin,
  Phys.\ Lett.\ B {\bf 272}, 39 (1991)
  [Erratum-ibid.\ B {\bf 319}, 545 (1993)]
  [hep-th/9503230].

\bibitem{Peskin}
Michael E. Peskin, Daniel V. Schroeder,
``An Introduction to Quantum Field Theory'',
Westview press, 1995.





\bibitem{Dou:1997fg}
  D.~Dou and R.~Percacci,
   Class.\ Quant.\ Grav.\ {\bf 15}, 3449 (1998),
  hep-th/9707239.


\bibitem{Souma:1999at}
  W.~Souma,
   Prog.\ Theor.\ Phys.\  {\bf 102}, 181 (1999),
  hep-th/9907027.

\bibitem{Reuter:2001ag}
  M.~Reuter and F.~Saueressig,
   Phys.\ Rev.\  D{\bf 65}, 065016 (2002),
  hep-th/0110054.


\bibitem{Lauscher:2001rz}
  O.~Lauscher and M.~Reuter,
   Class.\ Quant.\ Grav.\   {\bf 19}, 483 (2002),
  hep-th/0110021.


\bibitem{Litim:2001up}
  D.~F.~Litim,
   Phys.\ Rev.\ D{\bf 64}, 105007 (2001),
  hep-th/0103195.


\bibitem{Reuter:2003ca}
  M.~Reuter and H.~Weyer,
   Phys.\ Rev.\  D{\bf 69}, 104022 (2004),
  hep-th/0311196.

\bibitem{Reuter:2004nx}
  M.~Reuter and H.~Weyer,
   JCAP{\bf 0412}, 001 (2004),
  hep-th/0410119.



\bibitem{Codello:2007bd}
  A.~Codello, R.~Percacci and C.~Rahmede,
   Int.\ J.\ Mod.\ Phys.\  A{\bf 23}, 143 (2008),
  arXiv:0705.1769.



\bibitem{Benedetti:2009rx}
  D.~Benedetti, P.~F.~Machado and F.~Saueressig,
   Mod.\ Phys.\ Lett.\ A{\bf 24}, 2233 (2009),
  arXiv:0901.2984;
  Nucl.\ Phys.\ B{\bf 824}, 168 (2010),
  arXiv:0902.4630.



\bibitem{Niedermaier:2009zz}
  M.~R.~Niedermaier,
   Phys.\ Rev.\ Lett.\ {\bf 103}, 101303 (2009).


\bibitem{Benedetti:2012dx}
  D.~Benedetti and F.~Caravelli,
   JHEP{\bf 1206}, 017 (2012)
  [Erratum-ibid.\  {\bf 1210}, 157 (2012)],
  arXiv:1204.3541.

\bibitem{Dietz:2012ic}
  J.~A.~Dietz and T.~R.~Morris,
   JHEP{\bf 1301}, 108 (2013),
  arXiv:1211.0955.

\bibitem{Nagy:2012rn}
  S.~Nagy, J.~Krizsan and K.~Sailer,
   JHEP{\bf 1207}, 102 (2012),
  arXiv:1203.6564.

\bibitem{Christiansen:2012rx}
  N.~Christiansen, D.~F.~Litim, J.~M.~Pawlowski and A.~Rodigast,
  arXiv:1209.4038.

\bibitem{Demmel:2012ub}
  M.~Demmel, F.~Saueressig and O.~Zanusso,
   JHEP{\bf 1211}, 131 (2012),
  arXiv:1208.2038.

\bibitem{Codello:2013wxa}
  A.~Codello,
  arXiv:1304.2059.

\bibitem{Harst:2011zx}
  U.~Harst and M.~Reuter,
  JHEP {\bf 1105}, 119 (2011)
  [arXiv:1101.6007 [hep-th]].



\bibitem{Eichhorn:2009ah}
  A.~Eichhorn, H.~Gies and M.~M.~Scherer,
   Phys.\ Rev.\ D{\bf 80}, 104003 (2009),
  arXiv:0907.1828;
  K.~Groh and F.~Saueressig,
   J.\ Phys.\ A{\bf 43}, 365403 (2010),
  arXiv:1001.5032;
  A.~Eichhorn and H.~Gies,
   Phys.\ Rev.\ D{\bf 81}, 104010 (2010),
  arXiv:1001.5033.
  A.~Eichhorn,
   Phys.\ Rev.\ D{\bf 87} 124016 (2013),
  arXiv:1301.0632.


\bibitem{Christiansen:2014raa}
  N.~Christiansen, B.~Knorr, J.~M.~Pawlowski and A.~Rodigast,
  arXiv:1403.1232 [hep-th].

\bibitem{Koch:tobe}
   B.~Koch and P.~Rioseco,
``New black hole solutions for the improved Einstein-Hilbert-Maxwell action'',
work in progress;\\
   C.~Contreras, B.~Koch and P.~Rioseco,
 ``Coupling flows due to black hole solutions of the Einstein-Hilbert-Maxwell action'',
work in progress.


\bibitem{Frolov:2011ys}
  A.~V.~Frolov and J.-Q.~Guo,
  arXiv:1101.4995.

\bibitem{Bonanno:2012jy}
  A.~Bonanno,
   Phys.\ Rev.\ D{\bf 85}, 081503 (2012),
  arXiv:1203.1962.

\bibitem{Hindmarsh:2012rc}
  M.~Hindmarsh and I.~D.~Saltas,
   Phys.\ Rev.\ D{\bf 86}, 064029 (2012),
  arXiv:1203.3957.

\bibitem{Copeland:2013vva}
  E.~J.~Copeland, C.~Rahmede and I.~D.~Saltas,
  arXiv:1311.0881.

\bibitem{Stelle:1976gc} 
  K.~S.~Stelle,
     Phys.\ Rev.\ D {\bf 16}, 953 (1977).


\end{thebibliography}

\pagebreak

\begin{appendix}

\end{appendix}




\end{document}